\documentclass[12pt, a4paper]{article}

\usepackage[applemac]{inputenc}        % pour Mac
\usepackage[T1]{fontenc}
\usepackage[english,francais]{babel}
\usepackage{graphicx}
\usepackage{bbm}
\let\mathbb\relax \usepackage{bbold}

\newcommand{\pure}{{(0)}} 
 
\newcommand{\cst}{{(1)}} 
\newcommand{\acc}{{(1)}} 
 
\newcommand{\modi}{{(2)}} 
 
\newcommand{\dec}{{(d)}}

\newcommand{\eqw}{\sim}

\newcommand{\id}{\mathit{id}}

\newcommand{\inl}{\mathit{inl}} 
\newcommand{\inr}{\mathit{inr}}

\newcommand{\copair}[1]{[ #1 ]}

\newcommand{\copa}{[\;]}

\newcommand{\empt}{\mathbb{0}}

\newcommand{\tagg}{\mathtt{tag}}
\newcommand{\untag}{\mathtt{untag}}
\newcommand{\throw}{\mathtt{throw}}
\newcommand{\try}{\mathtt{try}}
\newcommand{\catch}{\mathtt{catch}}

\begin{document}

\thispagestyle{empty}  
\small
\title{\textbf{\large Program certification with computational effects}}

\author{\small J.-G.~Dumas$^\ast$, D.~Duval$^\ast$, \textbf{B.~Ekici}$^\ast$, D.~Pous$^{\dag}$\\
\small$^\ast$LJK, Universit\'{e} de Grenoble, France \\
\small$^{\dag}$LIP, ENS Lyon, France\\
\small\texttt{\{Jean-Guillaume.Dumas,Dominique.Duval,Burak.Ekici\}@imag.fr}\\
\small\texttt{Damien.Pous@ens-lyon.fr}}
\date{\small 8 octobre 2014}
\maketitle
Dynamic evaluation is a paradigm in computer algebra
which was introduced for computing with algebraic numbers. 
In linear algebra, for instance, dynamic evaluation can be used to apply 
programs which have been written for matrices with coefficients 
modulo some prime number 
to matrices with coefficients modulo some composite number. 
A way to implement dynamic evaluation in modern computing languages is 
to use the \textit{exceptions} mechanism provided by the language. 
In this paper, we pesent a proof system for exceptions which involves 
both raising and handling, by extending Moggi\rq{}s approach based on monads. 
Moreover, the core part of this proof system is dual to a proof system for 
the \textit{state effect} in imperative languages, which relies on the 
categorical notion of comonad \cite{DDFR12}. Both proof systems are 
implemented in the Coq proof assistant, and they are combined in order to deal 
with both effects at the same time.

The \textit{decorated logic} provides a rigorous formalism 
for proving properties of programs involving computational effects. 
To start with, let us describe the main features of the 
\textit{decorated logic for exceptions}.
Its syntax is given as follows, where $T$ is any exception name.
\renewcommand{\arraystretch}{1.20}
$$\begin{array}{lrcl}
\quad \textrm{Types: } & t &::=&
 A\mid B\mid \dots \mid
 t+t\mid \empt \mid V_T \\
 %t+t\mid\empt  \\
\quad \textrm{Terms: } & f &::=& \id \mid f\circ f\mid  
 \copair{f|f} \mid \inl \mid \inr \mid  \copa \mid \tagg_T \mid  \untag_T \\  
%\mid
% \tagg_T \colon V_T \to \empt \mid 
%  \untag_T \colon \empt \to V_T \\  %\mid
 %\copair{f|f}\mid \\ & & \copr_{t,t,1}\mid\copr_{t,t,2}\mid\copa_t \quad \\
\quad \textrm{Decorations: } & \dec &::=&\pure \mid \cst \mid \modi \\ 
\quad \textrm{Equations: } & e &::=& f\equiv f  \mid f\eqw f  
\end{array}$$
Here, $\empt$ is the empty type while $V_T$ represents the set of 
values which can be used as arguments for the exceptions with name $T$. 
Terms represent functions;
they are closed under composition and ``copairs'' (or case distinction), 
$\inl$ and $\inr$ represent the canonical inclusions 
into a coproduct (or disjoint union). 
The basic functions for dealing with exceptions are 
$\tagg_T\colon V_T \to \empt$ and $\untag_T\colon \empt \to V_T$.
A fundamental feature of the mechanism of exceptions is the distinction 
between \textit{ordinary} (or \textit{non-exceptional}) values
and \textit{exceptions}. 
While $\tagg_T$ encapsulates its argument (which is an ordinary value)  
into an exception, $\untag_T$ is applied to an exception for recovering 
this argument.
The usual $\throw$ and $\try/\catch$ constructions are built from 
the more basic $\tagg_T$ and $\untag_T$ operations \cite{DDER14}.
We use \textit{decorations} on terms for expressing 
how they interact with the exceptions.
If a term is \textit{pure},
which means that it has nothing to do with exceptions, 
then it has decoration $(0)$;
in particular, $\id^\pure$, $\inl^\pure$ and $\inr^\pure$ are pure. 
We decorate \textit{throwers} with $(1)$ and \textit{catchers} with $(2)$;
clearly $\tagg_T^\acc$ is a thrower while $\untag_T^\modi$ is a catcher.
A thrower may throw exceptions and must propagate any given exception, 
while a catcher may recover from exceptions. 
Using decorations provides a new schema where term signatures are constructed 
without any occurrence of a ``type of exceptions''. 
Thus, signatures are kept close to the syntax. 
In addition, decorating terms gives us the flexibility to cope 
with more than one interpretation of the set of exceptions. 
This means that with such an approach, 
any proof in this decorated logic is valid 
for different implementations of the exceptions. 
Besides, we have two different kinds of equality between terms: 
two terms are \textit{weakly equal} if 
they have the same behavior on ordinary values but may show differences 
on exceptions, 
and they are \textit{strongly equal} if they have the same behavior 
on both ordinary values and exceptions. 
We respectively use $\sim$ and $\equiv$ symbols to denote weak 
and strong equalities.

This syntax is enriched with a set of \textit{rules} that are 
decorated versions of the rules for \textit{equational logic}.  
The \textit{equivalence} rules ensure that both weak and strong equalities 
are equivalence relations. 
The \textit{hierarchy} rules allow to consider 
any pure term as a thrower, any thrower as a catcher, 
and any weak equality as a strong one. 
The ``copair'' construction $\copair{f, g}$ cannot be used when both 
$f$ and $g$ are catchers, since this would lead to a conflict when the 
argument is an exception. But $\copair{f, g}$ can be used when only $g$
is a catcher, it is the catcher $\copair{f,g}^\modi$
which is characterized by the equations 
$\copair{f,g} \circ inl \sim f$ and $\copair{f,g} \circ inr \equiv g$. 
This means that exceptional arguments  
are treated by $\copair{f, g}$ as they would be by $g$. 
The \textit{substitution} rule for weak equations 
$f_1^\modi \sim f_2^\modi \Longrightarrow f_1 \circ g \sim f_2 \circ g$ 
is valid \textit{only} when the substituted term $g$ is \textit{pure}. 
The behaviour of the $\untag_T$ functions is given by the rules 
$\untag_T \circ \tagg_T \sim \id_T$ and 
$\untag_T \circ \tagg_R  \sim \copa_R \circ \tagg_T$ for all exception names 
$T \neq R$ (where $\copa_R:\empt\to R$ is the canonical embedding). 

Such a formal system enables us to prove properties of programs 
involving exceptions. 
The decorated logic for states and the decorated logic for exceptions, 
which are mutually dual, are implemented in Coq \cite{DDER14}.
For instance, we have used these logics for proving 
the primitive properties of the state effect proposed in \cite{PP02}
and the dual properties of exceptions.
To cope with programs including both states and exceptions at the same time, 
we have composed these Coq implementations, 
by merging the syntax and the rules.  
We have also translated the basic imperative programming language IMP 
in our library, as well as the language IMP\_EXC made of IMP 
extended with exceptions. 
We have used this implementation to prove some properties of IMP and 
IMP\_EXC programs. For instance, we have checked some simple properties 
of programs calculating the rank of a (2x2) matrix modulo a composite number 
using dynamic evaluation \cite{DDER14}.  

We would like to be able to prove more general properties of algorithms 
for linear algebra using dynamic evaluation implemented through exceptions. 
For this purpose, we plan to implement Hoare logic for IMP\_EXC 
in decorated terms. We also plan to study other effects 
(partiality, IO, non-determinism, \ldots) and to compose them 
in a systematic way.

\end{document}